\documentclass[]{IEEEtran}

\usepackage{amsmath}
\usepackage{amssymb}
\usepackage{graphicx}
\graphicspath{{figures/}}
\newcommand{\micron}{$\mu$m}

\begin{document}

\title{Two-dimensional subwavelength focusing using a slotted meta-screen}

\author{Lo\"{i}c Markley, \IEEEmembership{Student Member, IEEE} and George V. Eleftheriades, \IEEEmembership{Fellow, IEEE}%
\thanks{The authors are with the Edward S. Rogers Sr. Department of Electrical and Computer Engineering, University of Toronto, 40 St. George Street, Toronto, Ontario, M5S 3G4, Canada (e-mail: gelefth@waves.utoronto.ca).}%
\thanks{Color versions of one or more of the figures in this letter are available online at http://ieeexplore.ieee.org.}}

\maketitle

\begin{abstract}
A transmission screen with subwavelength slots is described which can focus electromagnetic radiation to a two-dimensional subwavelength spot.  Unlike negative-refractive-index focusing implementations, this ``meta-screen" does not suffer from image degradation when losses are introduced and is easily scalable from microwave to Terahertz frequencies and beyond.  The slotted geometry is designed using a theory of shifted beams to determine the necessary weighting factors for each slot element, which are then converted to appropriate slot dimensions.  The screen was designed using a pair of rotated orthogonal focusing axes to produce a spot on a focal plane located $0.25\lambda$ away.  Full-wave simulations are corroborated by experimental measurements at 10 GHz reporting a full-width half-maximum beam width of $0.27\lambda$ along the $x$-axis and $0.38\lambda$ along the $z$-axis.
\end{abstract}



\section{Introduction}

The biggest obstacle to developing high-resolution imaging systems has been the diffraction limit.  Electromagnetic fields scattering off an object are composed of a rich spectrum of propagation vectors.  The high transverse spatial frequency components corresponding to the finer details of the object cannot propagate very far into space and attenuate rapidly.  The loss of these evanescent waves outside the near field of the object leads to a loss in image resolution of details below the order of a wavelength.  The field of near-field optical microscopy emerged in 1972 to overcome this effect by placing tiny apertures extremely close to the object to be imaged \cite{AshNicholls:1972,PohlDenkLanz:1984}.  This exploited the strong contribution of evanescent field components, but high attenuation and precise probe placements increased the cost and complexity of this technique.  When J. B. Pendry proposed the ``perfect lens" in 2000, the field of negative-refractive-index metamaterials provided an alternative method to achieve subwavelength focusing through the amplification of evanescent field components \cite{Pendry:2000}.  Although this technique was successfully demonstrated in planar form \cite{GrbicEleftheriades:2004} and in the quasi-static limit \cite{FangLeeSunZhang:2005,MesaFreireMarquesBaena:2005,Wiltshire:2007}, the presence of loss heavily degraded the image quality.  An alternative approach was through the plane-wave expansion of aperture fields with direct integration leading to near-field plate implementations \cite{GrbicJiangMerlin:2008} and holographic concepts leading to transmission screens \cite{EleftheriadesWong:2008}.  A conceptually different way to synthesize such structures using the principle of spatially shifted beams was introduced in \cite{MarkleyWongWangEleftheriades:2008} for one-dimensional focusing.  In this Letter, the shifted-beam approach is extended to two-dimensions, leading to the development of a meta-screen that can focus an incident wave to a two-dimensional spot.  With the focal plane pushed back to a distance of $0.25\lambda$ away from the screen, this structure shows a marked improvement over the one-dimensional focusing structures presented in \cite{GrbicJiangMerlin:2008,EleftheriadesWong:2008,MarkleyWongWangEleftheriades:2008}.  Composed of an array of slot antennas, the screen does not suffer from focal degradation due to losses and is scalable from microwave to Terahertz frequencies and beyond.

\begin{figure}
    \centering
    \includegraphics[clip,scale=1]{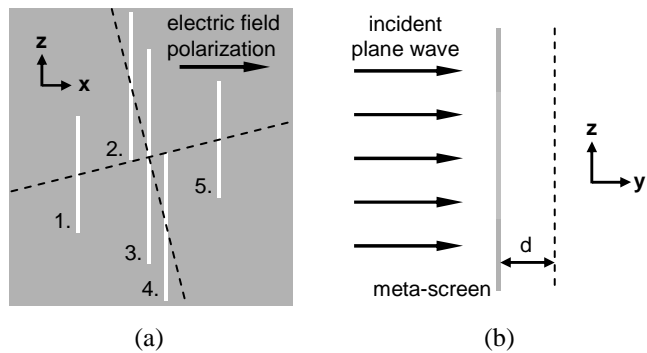}
    \caption{(a) The geometry of the slotted meta-screen showing the numbering of the slots from one to five.  The slots are arranged along orthogonal focusing axes with their centers equidistant from the origin to provide two-dimensional focusing.  (b) The meta-screen orientation with respect to the incident field.  The electric field is polarized along the x-direction with the focal plane parallel to the screen but separated by a distance $d$.}
    \label{fig:diagrams}
\end{figure}


\section{Theoretical Development}

The field pattern produced by an array of elements can be analyzed by superposing the field patterns of each individual element.  In the far field, this pattern can be separated into the pattern of an identical element located at the origin together with a phase correction factor that accounts for the element's actual position.  In the near field, however, this ``array factor" does not appear, and the field distribution is not shifted in phase but in space.  This near-field property of spatially shifted beams can be utilized to synthesize field distributions such as subwavelength focal spots by appropriately weighting the shifted beams produced by closely spaced array elements and considering their superposition.

\begin{subequations}
    \begin{eqnarray}
        & & E_x\Big|_{\textrm{ff}} = E_x(x,y,z) e^{-j k x_0 \sin{\theta}\cos{\phi}} e^{-j k z_0 \cos{\theta}} \label{eqn:Exshiftedff} \\
        & & E_x\Big|_{\textrm{nf}} = E_x(x-x_0,y,z-z_0) \label{eqn:Exshiftednf}
    \end{eqnarray}
\end{subequations}

If $E_x(x,y,z)$ is the transverse electric field produced by a single $z$-directed slot element at the origin, then the fields in the near- and far-field regions produced by a slot shifted to a new location on the $xz$-plane $(x_0,z_0)$ would be given by \eqref{eqn:Exshiftedff} and \eqref{eqn:Exshiftednf}, respectively.  Superposing the near-field contributions from $N$ elements, each located at $(x_n,z_n)$ and weighted by a complex constant factor $w_n$, produces a total field described by \eqref{eqn:Extotal}.  The vector representation condenses the fields from each element into an $N$-element row vector and the weights associated with each slot into an $N$-element column vector.

\begin{eqnarray} \label{eqn:Extotal}
    E_{x,total}(x,z) & = & \sum_{n=1}^N w_n E_x(x-x_n,y,z-z_n)  \nonumber \\
& = & \sum_{n=1}^N w_n E_{x,n}(x,z) = \mathbf{E_x} \cdot \mathbf{w}
\end{eqnarray}

Given a target field distribution $f(x,z)$, the objective is to choose optimal weights $w_{opt}$ such that $\mathbf{E_x}(x,z) \cdot \mathbf{w_{opt}} \approx f(x,z)$.  By multiplying both sides of the approximation by $E^*_{x,m}(x,z)$ and integrating over $x$ and $z$ for all $m$, a set of $N$ independent equations is created.  This can be recognized as a method of moments field expansion using the individual element current distributions as basis functions and field patterns as testing functions.  Discretizing $E_{x,n}(x,z)$ and $f(x,z)$ over $x$ and $z$ and writing them as the column vectors in $\mathbf{A}$ and $\mathbf{b}$, respectively, the optimal weights can be calculated using \eqref{eqn:leastsquares}.

\begin{equation} \label{eqn:leastsquares}
        \mathbf{w_{opt}} = \left(\mathbf{A^H}\mathbf{A}\right)^{-1}\mathbf{A^H} \cdot \mathbf{b}
\end{equation}

One-dimensional focusing has been demonstrated using an array of three slots spaced by $\lambda/10$ \cite{MarkleyWongWangEleftheriades:2008} but a direct extension to two-dimensions could not be applied without causing the half wavelength slots to overlap longitudinally.  By rotating the focal axes slightly, however, as illustrated in Fig. \ref{fig:diagrams}a, a geometry was found that still permits tight slot spacings.  The centers of the four satellite slots are located in the $xz$-plane at $(-0.20\lambda,-0.05\lambda)$, $(-0.05\lambda,0.20\lambda)$, $(0.05\lambda,-0.20\lambda)$, and $(0.20\lambda,0.05\lambda)$.


\section{Full-wave Simulations}

The optimal weights were calculated using \eqref{eqn:leastsquares} for a focal plane $0.25\lambda$ away from the screen and a target Gaussian distribution with a full-width half-maximum (FWHM) beam width of $0.2\lambda$.  Normalized with respect to the central slot weighting factor, the weights of the satellite slots were calculated to be $w_1 = w_5 = 0.271 \angle 170.2^\circ$ for the slots along the $x$-directed focal axis and $w_2 = w_4 = 0.306 \angle 173.5^\circ$ for the slots along the $z$-directed focal axis.  Full-wave simulations using Ansoft's HFSS of very thin slots fed by voltage sources confirmed the theoretical results.

\begin{figure}
    \centering
    \includegraphics[clip,scale=1]{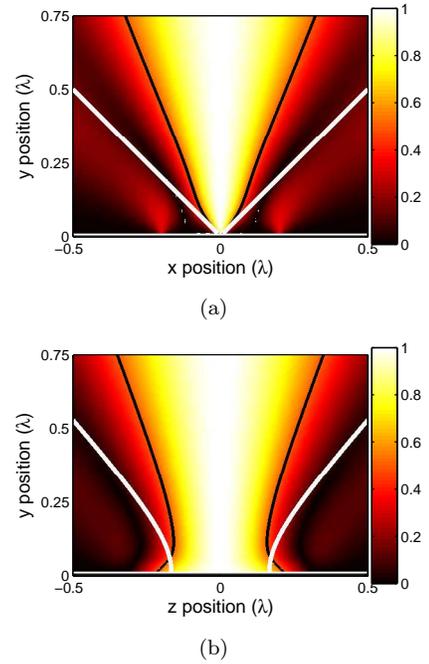}
    \caption{Full-wave simulations of the electric field magnitude emanating from the meta-screen under plane wave excitation.  The fields are normalized transverse to the $y$-axis at every $y$ position to highlight the profile of the beam as it progresses away from the screen.  The $xy$-plane is shown in (a) with the $zy$-plane in (b).  The thinner black contour indicates the FWHM beam width of the focused beam while the thick white contour indicates the FWHM beam width of a single slot.}
    \label{fig:sections}
\end{figure}

To perform experimental verification of the focusing array, plane-wave incidence was considered.  The alternating signs of the weights were implemented by lengthening the central slot to make it capacitive and shortening the satellite slots to make them inductive.  The mutual coupling between the closely spaced slots caused them to resonate strongly and contributed to the lowered amplitude of the fields produced from the satellite slots.  The complete $5 \times 5$ admittance matrix was calculated using self and mutual admittance equations \cite{EleftheriadesRebeiz:1993} for a variety of slot lengths while the induced slot currents were determined by integrating the assumed magnetic current distribution over the incident field.  The lengths corresponding closely to the optimal weights were optimized through simulations using the Multiradius Bridge Current software that implements a thin-wire formulation of the electromagnetic moment method \cite{TilstonBalmain:1990}.  The final set of lengths was determined to be $L_1 = L_5 = 0.330\lambda$, $L_2 = L_4 = 0.415\lambda$, and $L_3 = 0.6075\lambda$ with the widths all set to $0.01\lambda$.  These dimensions were successfully verified in HFSS to produce a subwavelength focal spot.  The field profile as it progresses away from the screen is shown in Fig. \ref{fig:sections}.

\section{Experimental Results}

\begin{figure}
    \centering
    \includegraphics[clip,scale=1]{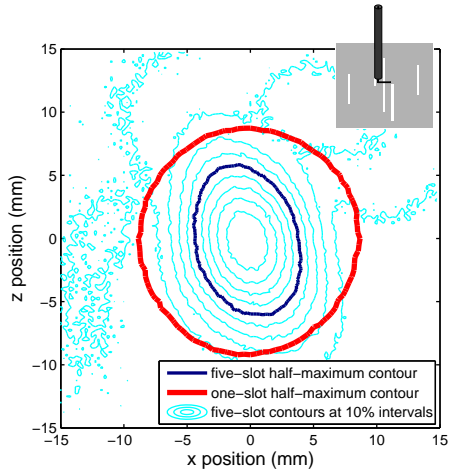}
    \caption{Experimental measurements of the electric field at 10 GHz over a plane 7.5 mm away from the screen.  The thin blue contours are spaced at 10\% intervals from 10\% to 90\% of the peak magnitude while the thicker blue oval contour represents the FWHM focal spot.  The thick red round contour represents the FWHM focal spot of the diffraction limited beam produced by a single slot.  Asymmetries in the measured field profile were exacerbated by the asymmetric coupling between the probe antenna and the slots (see inset).}
    \label{fig:measurements}
\end{figure}

The meta-screen was fabricated from a sheet of 2 mil stainless steel cut using a laser with a cutting accuracy of 5 \micron.  At 10 GHz, the slot dimensions are 9.9 mm $\times$ 0.3 mm, 12.45 mm $\times$ 0.3 mm, and 18.225 mm $\times$ 0.3 mm with the centers of the satellite slots spaced 6 mm $\times$ 1.5 mm away from the center of the central slot.  A plane wave excitation was produced by collimating the beam from an X-band horn antenna through a Rexolite biconvex lens and placing the screen at the Gaussian beam waist.  An Agilent E8364B network analyzer was used to collect the field data by recording the transmission scattering parameters between the horn antenna and a co-axial cable probe whose exposed inner conductor was bent 90 degrees to form a miniature dipole receiver antenna (see Fig. \ref{fig:measurements} inset).  A high-precision custom built XYZ-stage from Newmark Systems positioned the probe to collect the transverse electric field data on a plane 7.5 mm above the surface of the screen.

\begin{figure}
    \centering
    \includegraphics[clip,scale=1]{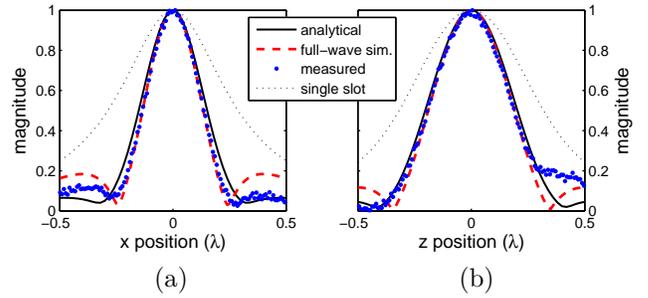}
    \caption{A comparison of the field profiles along the $x$-axis (a) and the $z$-axis (b).  Measured data is plotted alongside full-wave simulation data and analytical predictions, with the diffraction limited beam from a single slot given for comparison.  The focal plane of the analytic and simulation data was moved to $0.28\lambda$ so the single-slot data would be consistent with the measurements.}
    \label{fig:comparisons}
\end{figure}

The meta-screen field measurements were compared to the field produced by a single slot, which has broadened to the diffraction limit at a distance of $\lambda/4$, and to the background field measured with the screen removed.  The peak field magnitude from the single slot was 7.8 dB lower than the background field while the focused beam peak magnitude was 15.4 dB lower than the background field.  The focused FWHM beam widths relative to the single-slot beam widths were measured to be 0.463 along the $x$-axis ($0.271\lambda$) and 0.643 along the $z$-axis ($0.385\lambda$).  Fig. \ref{fig:comparisons} presents the analytical predictions together with the simulation results and experimental measurements.  The close agreement between the simulation data, in which losses were not taken into account, and the measurement data indicates that the performance of the meta-screen is not affected by the introduction of losses.


\section{Conclusion}

In summary, a meta-screen was designed to produce a two-dimensional subwavelength focal spot $0.25\lambda$ away from the screen in response to a plane wave excitation.  A shifted-beam approach to near-field focusing was used to develop the structure out of an array of closely spaced slot antennas.  The element field patterns were weighted out-of-phase by changing the lengths of the slots so they would resonate together.  Full-wave simulations were performed to verify the focusing behavior followed by confirmation through experimental measurements at 10 GHz.  The advantages of the meta-screen over alternative implementations relate primarily to its simplicity, which allows it to be easily scaled in frequency and to be resistant to focal degradation in the presence of losses.

\begin{thebibliography}{10}
\providecommand{\url}[1]{#1}
\csname url@samestyle\endcsname
\providecommand{\newblock}{\relax}
\providecommand{\bibinfo}[2]{#2}
\providecommand{\BIBentrySTDinterwordspacing}{\spaceskip=0pt\relax}
\providecommand{\BIBentryALTinterwordstretchfactor}{4}
\providecommand{\BIBentryALTinterwordspacing}{\spaceskip=\fontdimen2\font plus
\BIBentryALTinterwordstretchfactor\fontdimen3\font minus
  \fontdimen4\font\relax}
\providecommand{\BIBforeignlanguage}[2]{{%
\expandafter\ifx\csname l@#1\endcsname\relax
\typeout{** WARNING: IEEEtran.bst: No hyphenation pattern has been}%
\typeout{** loaded for the language `#1'. Using the pattern for}%
\typeout{** the default language instead.}%
\else
\language=\csname l@#1\endcsname
\fi
#2}}
\providecommand{\BIBdecl}{\relax}
\BIBdecl

\bibitem{AshNicholls:1972}
E.~A. Ash and G.~Nicholls, ``Super-resolution aperture scanning microscope,''
  \emph{Nature}, vol. 237, pp. 510--512, Jun. 1972.

\bibitem{PohlDenkLanz:1984}
D.~W. Pohl, W.~Denk, and M.~Lanz, ``Optical stethoscopy: image recording with
  resolution $\lambda/20$,'' \emph{Appl. Phys. Lett.}, vol.~44, no.~7, pp.
  651--653, Apr. 1984.

\bibitem{Pendry:2000}
J.~B. Pendry, ``Negative refraction makes a perfect lens,'' \emph{Phys. Rev.
  Lett.}, vol.~85, no.~18, pp. 3966--3969, Oct. 2000.

\bibitem{GrbicEleftheriades:2004}
A.~Grbic and G.~V. Eleftheriades, ``Overcoming the diffraction limit with a
  planar left-handed transmission-line lens,'' \emph{Phys. Rev. Lett.},
  vol.~92, no.~11, p. 117403, Mar. 2004.

\bibitem{FangLeeSunZhang:2005}
N.~Fang, H.~Lee, C.~Sun, and X.~Zhang, ``Sub–diffraction-limited optical
  imaging with a silver superlens,'' \emph{Science}, vol. 308, pp. 534--537,
  Apr. 2005.

\bibitem{MesaFreireMarquesBaena:2005}
F.~Mesa, M.~J. Freire, R.~Marqu\'{e}s, and J.~D. Baena, ``Three-dimensional
  superresolution in metamaterial slab lenses: experiment and theory,''
  \emph{Phys. Rev. B}, vol.~72, p. 235117, Dec. 2005.

\bibitem{Wiltshire:2007}
M.~C.~K. Wiltshire, ``Radio frequency ({RF}) metamaterials,'' \emph{Phys. Stat.
  Sol. (B)}, vol. 244, no.~4, pp. 1227--1236, Mar. 2007.

\bibitem{GrbicJiangMerlin:2008}
A.~Grbic, L.~Jiang, and R.~Merlin, ``Near-field plates: subdiffraction focusing
  with patterned surfaces,'' \emph{Science}, vol. 320, pp. 511--513, Apr. 2008.

\bibitem{EleftheriadesWong:2008}
G.~V. Eleftheriades and A.~M.~H. Wong, ``Holography-inspired screens for
  sub-wavelength focusing in the near field,'' \emph{{IEEE} Microwave Wireless
  Compon. Lett.}, vol.~18, no.~4, pp. 236--238, Apr. 2008.

\bibitem{MarkleyWongWangEleftheriades:2008}
L.~Markley, A.~M.~H. Wong, Y.~Wang, and G.~V. Eleftheriades, ``A spatially
  shifted beam approach to subwavelength focusing,'' \emph{Phys. Rev. Lett.} (in
  press).

\bibitem{EleftheriadesRebeiz:1993}
G.~V. Eleftheriades and G.~M. Rebeiz, ``Self and mutual admittance of slot
  antennas on a dielectric half-space,'' \emph{International Journal of
  Infrared and Millimeter Waves}, vol.~14, no.~10, pp. 1925--1946, 1993.

\bibitem{TilstonBalmain:1990}
M.~A. Tilston and K.~G. Balmain, ``A multiradius, reciprocal implementation of
  the thin-wire moment method,'' \emph{{IEEE} Trans. Antennas Propagat.},
  vol.~38, no.~10, pp. 1636--1644, Oct. 1990.

\end{thebibliography}
\end{document}